\documentclass[11pt,letterpaper]{article} %% two column, final layout
\usepackage[english]{babel}
\usepackage{ol2}
\usepackage[draft,implicit=false]{hyperref}
\usepackage{amsmath}
\usepackage{amsfonts}
\usepackage{graphicx}
\usepackage{caption}
\usepackage{subcaption}
\usepackage{varioref}
\usepackage{cite}

\RequirePackage{snapshot}
\begin{document}

\title{Stable Planar Mesoscopic Photonic Crystal Cavities}

%% For REVTeX it is possible to automate superscript and e-mail callouts with the superscriptaddress option; see REVTeX4 documentation.

\author{G.~Magno,$^{1,3,*}$ A.~Monmayrant,$^{1,2}$ M.~Grande,$^{3}$ F.~Lozes-Dupuy,$^{1,2}$ O.~Gauthier-Lafaye,$^{1,2}$ G.~Cal\`{o},$^{3}$ V.~Petruzzelli$^{3}$}

\address{
$^1$CNRS, LAAS, 7 avenue du colonel Roche, F-31400 Toulouse, France \\
$^2$Univ. de Toulouse, LAAS, F-31400 Toulouse, France \\
$^3$Dipartimento di Ingegneria Elettrica e dell'Informazione (DEI), Via Re David 200, Politecnico di Bari 70125 (Italy)\\
$^*$Corresponding author: giovanni.magno@poliba.it
}

\begin{abstract}Mesoscopic self-collimation in mesoscopic photonic crystals with high reflectivity is exploited to realize a novel high-Q factor cavity by means of  mesoscopic PhC planar mirrors. These mirrors efficiently confine a mode inside a planar Fabry-Perot-like cavity, due to a beam focusing effect that stabilises the cavity even for small beam sizes, resembling the focusing behaviour of curved mirrors. Moreover, they show an improved reflectivity with respect to their standard distributed Bragg reflector counterparts that allows higher compactness.  A Q factor higher than $10^4$ has been achieved for an optimized 5-period-long mirror cavity. The optimization of the Q factor and the performances in terms of energy storage, field enhancement and confinement are detailed.\end{abstract}

%\begin{abstract}Mesoscopic self-collimation in mesoscopic photonic crystals with high reflectivity is exploited to realize a novel high-Q factor stable planar cavity by means of planar mirrors. These mesoscopic PhC planar mirrors are able to efficiently confine a mode inside Fabry-Perot-like cavity, resembling the focusing behaviour of curved mirrors. Moreover, they show an improved reflectivity with respect to their standard distributed Bragg reflector counterparts that allows higher compactness.  A Q factor higher than $10^4$ has been achieved for an optimized 5-period-long mirror cavity. The optimization of the Q factor and the performances in terms of energy storage, field enhancement and confinement are detailed.\end{abstract}

\noindent Since its demonstration in 1999 \cite{Kosaka1999} self-collimation (SC) has attracted a lot of efforts. In particular, it allows diffraction-less propagation of light within photonic crystals (PhCs) without any typical confining mechanism - as index or band-gap guiding - allowing beam crossing  without any crosstalk \cite{Chigrin2002}. Mesoscopic self-collimation (MSC), as recently proposed in mesoscopic PhCs (MPCs) \cite{Arlandis2012}, extends the paradigm of SC allowing its coexistence with many other interesting optical properties as tailored overall reflectivity or slowlight. Reflectivity controlled  MPCs elegantly solve the problem of the  input interface impedance mismatch without affecting the feasibility and the simplicity of the structure\cite{Magno2014}. 
Indeed, in PhCs, the impedance mismatch arising at the input interface has been one troublesome issue. Several solutions have been proposed, as graded PhC \cite{Chuang2010,Shen2011} or half-holes \cite{Arlandis2012,Mocella2009,Rakich2006}. However, these solutions are difficult to implement in real structures. 
Moreover, MSC allows the conception of two basic building blocks, the anti-reflection (AR) and the high-reflection (HR) MPCs, that can be exploited to design more complex optical devices\cite{Magno2014}. In this letter, we show how it is possible to exploit the HR MPC as a very efficient mirror to conceive a novel class of planar stable microcavities based on the MSC effect.
\begin{figure}[b!]	
\centering
\includegraphics[width=0.60\columnwidth]{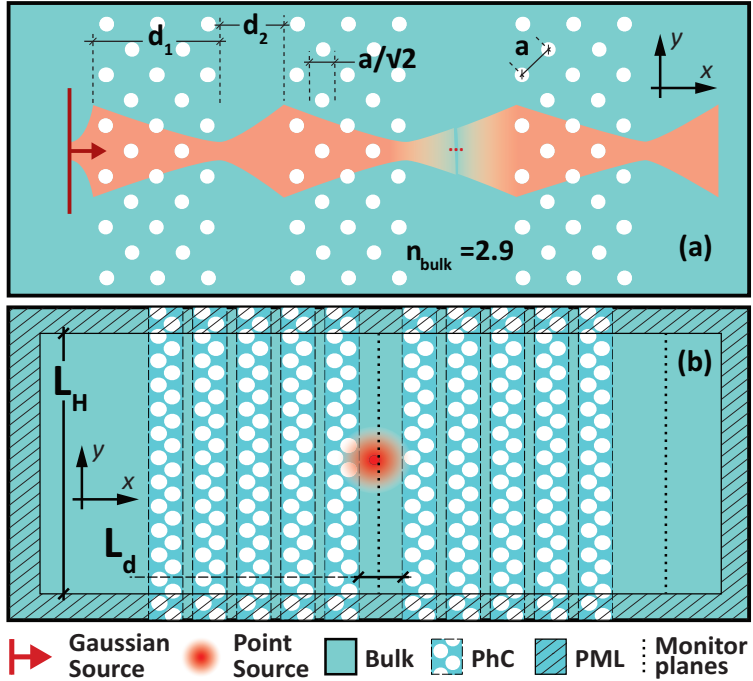}
\caption{Sketches (a) of the MPC mechanism, and (b) the computational cell.}
\label{Fig:geometry}
\end{figure}

 Fig.~\ref{Fig:geometry}(a)   shows a sketch of a MPC obtained by alternating slabs of length $d_1$ of a 45-degree-tilted plain PhC and slabs of length $d_2$ of bulk material ($n_{bulk}=2.9$ corresponding to the effective refractive index at $1.55 ~\mu m$ of the suspended GaAs membrane described in \cite{PTL08-20-24-2120-2122}). For sake of simplicity and to preserve the feasibility, the slabs of PhC consist of an integer number of rows $N_{row}$ of a square-lattice (lattice constant \textit{a}) of air hole (radius $r=0.28\,a$) etched into the same high-index material constituting the bulk slabs. This implies that $\left. d_1=N_{row}a/\sqrt{2} \right.$.
% As shown in \cite{Arlandis2012}, to obtain the MSC effect, a zero accumulated curvature in one mesoscopic period must be assured.%
MSC is obtained when the total spreading over one mesoscopic period averages to zero\cite{Arlandis2012}.
This condition relies on the \emph{curvature index} (see \cite{Arlandis2012, Magno2014} for a complete derivation) that governs the lateral spreading of the beam just like group index governs the temporal spreading of wavepacket.
Moreover, since this structure can be seen as a 1D super-crystal, in order to obtain an HR MPC, a multilayer reflector design rule has to be considered. Combining these two requirements, the following system must be solved:
\begin{equation}
\left\lbrace
\begin{array}{ll}
d_1/n_c(u)+d_2/n_{bulk}=0 \\
d_1n(u)=m\lambda/2, \; \; d_2n_{bulk}=p\lambda/2
\end{array}
\right.
\label{eq:HRdes}
\end{equation}
where ($m,p$) are both integers, $n_c(u)$ and $n(u)$ are the curvature index and the phase index, respectively, of the plain PhC \cite{Arlandis2012}, $n_{bulk}$ is both the phase and the curvature index in the bulk (identical values), $\lambda$ and $u=a/ \lambda$ are the wavelength and the normalised frequency of the propagating beam, respectively.
This system has a high number of solutions, which increases as $N_{row}$ increases. For sake of demonstration we focus on one solution of the Eq.~(\ref{eq:HRdes}) corresponding to the following parameters: $d_1=9.192\,a$ ($N_{row}=13$), $d_2=2.612\,a$, and central frequency $u_0=0.2311\,a/\lambda$ (see \cite{Magno2014}). 
The MSC model  has been developed in \cite{Arlandis2012} using gaussian beam formalism. Furthermore, since we expect the HR to be an efficient mirror able to reflect  the entire beam within few periods, the collimation of the reflected beam does not appear so crucial. However, we are interested in exploiting the structure as a mirror to confine non-paraxial spherical waves. This requires a much wider acceptance angle than a paraxial gaussian beam with a finite waist. Therefore, we want the HR structure to work in a regime which ensures a focusing behaviour in reflection. As shown in \cite{Arlandis2012}, this behaviour is achievable in a spectral range slightly above the frequency of self-collimation.

\begin{figure}[b!]	%\label{Fig:geometry}
\centering
\includegraphics[width=0.60\columnwidth]{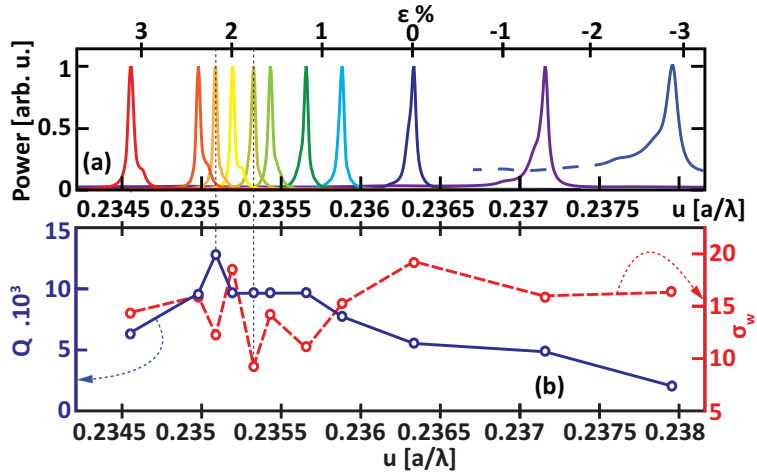}
\caption{(a) Self-normalised power spectra recorded in $x=0$. On the higher axis the corresponding value of the $\epsilon$ parameter is reported. (b) Calculated Q factor (blue solid curve and left axis) and $\sigma_W$ (red dashed curve and right axis) as a function of the normalised frequency.}
\label{Fig:Qfactor}
\end{figure}

Starting from this solution we can easily conceive a Fabry-Perot-like cavity by considering a certain number of mesoscopic periods placed at the sides of a defect realised by means of a bulk slab of length $L_d$. In this letter, we will consider a cavity with 5-mesoscopic-period-long mirrors. To test and inspect this novel cavity design, a well-established 2D-FDTD code \cite{Oskooi2010687} has been used. 
Thanks to the scale invariance of Maxwell's equations, in this letter we use dimensionless unit (by putting $\epsilon_0\,=\,\mu_0\,=\,c\,=\,1$) normalised to the PhC lattice period $a$, according to \cite{Oskooi2010687}.
The computational cell is depicted in Fig.~\ref{Fig:geometry}(b).
The vertical size of the cell is $L_\text{H}=100\,a$ and a resolution of $25$ points per $a$ has been used. In order to avoid reflections, the cell is surrounded by perfectly matched layers (PML). An omnidirectional point source is placed in a random position near the centre of the cavity, in order to excite all the possible supported modes. TE polarization is considered, with a $\vec{H} =H_z \vec{e_z}$, where $\vec{e_z}$ is the out-of-plane unitary vector. The electromagnetic fields have been recorded along two planes within the cell: inside the defect, for $x=0$, and  outside the cavity, for $x=75\,a$ (see dotted lines in Fig.~\ref{Fig:geometry}(b)).  A preliminary calculation performed on a single mirror shows the presence of a band-gap in the range $u\in[0.2320; 0.2389]$.  Thus, a first set of simulations has been performed on the whole cavity by means of a temporal gaussian source centred inside the band-gap and wide enough to spectrally cover it. To finely tune the cavity length we define:
\begin{equation}
L_d( \epsilon )= L_0 (1+ \epsilon), \; \; \epsilon \in [ -3 \%,3 \% ], 
\label{eq:Ld}
\end{equation}
where $L_0=2d_2$ and $\epsilon$ is the tuning parameter. All simulations last long enough to ensure a reduction of the stored energy by a factor $10^{3}$. By recording the temporal magnetic field $H_z$ on the plane inside the defect ($x=0$) and by using a zero padding technique, we are able to achieve power spectra with a resolution of 
$6.1 \cdot 10^{-6}\,a/\lambda$.
Several simulations, spanning the $\epsilon$ value from $-3 \%$ to $3 \%$, have been performed. Fig.~\ref{Fig:Qfactor}(a) shows the power spectra retrieved on the plane within the defect (normalised to the maximum of each resonant peak) for several values of the parameter $\epsilon$. For each peak the normalised frequency ($u_0$) corresponding to the maximum and the full width at half maximum ($\Delta u$) have been retrieved. Finally the Q factor has been estimated as follows: 
\begin{equation}
%Q=\frac{u_0}{\Delta u}.
Q=u_0/\Delta u.
\label{eq:Q}
\end{equation}
As can be seen in Fig.~\ref{Fig:Qfactor}(b), the highest Q factor of 12840 is obtained for $\epsilon=2.15 \%$ which corresponds to $u_\textit{cav}=0.2351$. It is worth underlining that the dependence of $u_0$ with $\epsilon$ is almost linear in the considered range.
Supposing that the best working regime corresponds to the best confinement of the beam inside the cavity, the waist of the confined beam has been calculated as a function of the parameter $\epsilon$. The confinement, subsequently, can be  estimated by means of a weighted standard deviation of the $y$ position of the beam. Assuming $W(y)={|H_z(0,y)|}^2$ as the weighting function, that is the square modulus of the Fourier transform of the $z$ component of the temporal magnetic field recorded on the plane inside the defect ($x=0$), it is possible to define the \textit{spreading} $\sigma_W$ of the beam inside the defect as follows:

\begin{equation}
\sigma_W=\sqrt{\frac{\sum_{y}{{(y-m_W)}^2}W(y)}{\sum_{y}{W(y)}}},
\label{eq:WeightedSigma}
\end{equation}
where $m_W$ represents the weighted mean defined as
\begin{equation}
m_W=\sum_{y}{\frac{yW(y)}{\sum_{y}{W(y)}}}.
\label{eq:WeightedMean}
\end{equation}

As can be seen in Fig.~\ref{Fig:Qfactor}(b), the minimum $\sigma_W$ is achieved for $\epsilon=1.75\%$ for the corresponding normalised frequency $u_\sigma=0.2353$. Thus, the highest confinement is achieved for a configuration slightly different from the one showing the highest Q factor, unveiling a trade-off between mode confinement and losses. However, it is worth noticing that the Q factor remains higher than 9500 for $\epsilon$ $\in [1.14 \%, 2.37 \%]$, showing a good confinement of the mode. Moreover, since $u_0$ is linear with respect to $\epsilon$, this design is easily tunable and robust to fabrication tolerances.

To better understand the mechanism underlying the behaviour of these planar cavities and to obtain a basis to compare the performances, an equivalent cavity obtained by the substitution of the PhC slabs by slabs of equivalent medium has been analysed.  At $u_\text{cav}=0.2351$ the planar PhC shows an effective refractive index $n_{\text{eff}}=2.4612$, that can be easily retrieved by means of the Plane Wave Expansion Method (PWEM). For an optimized equivalent cavity, of length $L_d=2d_2$, we have calculated a maximum Q factor of about 17. This value is about 755 times lower than the maximum Q factor observed for the MPC cavity, ascribable to lack of any lateral confining mechanism that forbids the establishment of a stable resonance. Even if the MPC cavity is compared with the 1D equivalent cavity, that accounts for the  lateral confining mechanism  provided by MSC, the maximum Q factor achieved is about 240 for $\epsilon=0 \%$. This value is about 53 times lower than the maximum Q factor observed for the MPC 
cavity. This is ascribable to an improved reflectivity of the MPC counterpart ($R>98\%$ for a 5-period-long mirror) due to the modal mismatch at the PhC-bulk interfaces existing between the Bloch mode inside the PhC slabs and the free propagating beam in the bulk slabs. It is worth pointing out that, in the 1D case, to achieve comparable reflectivity a much longer 17-period mirror is needed. Thus, MPC mirrors ensure higher compactness for the same reflectivity and higher Q factor values are expected for cavities having longer mirrors.

\begin{figure}[b!]
\centering
\includegraphics[width=0.60\columnwidth]{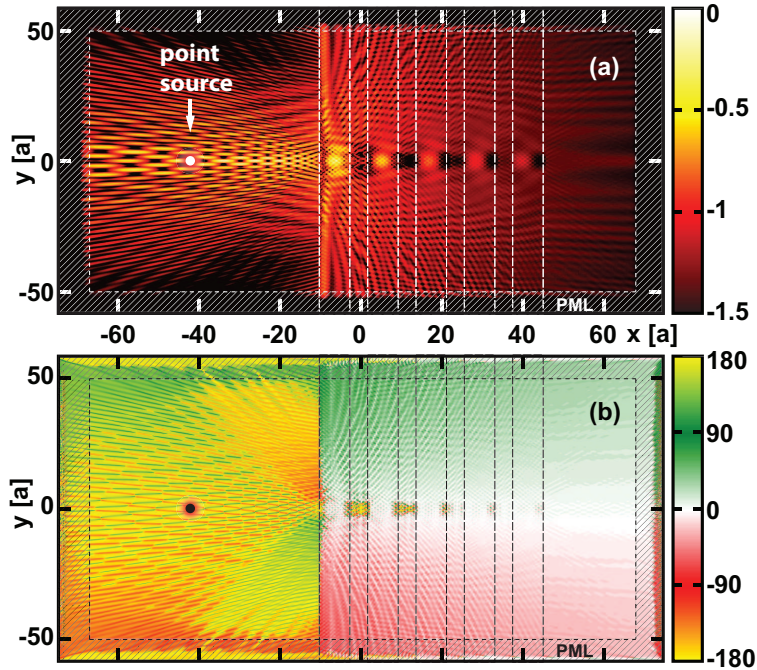}
\caption{(a) Logarithm of absolute value and (b) angle of the Poyinting vector of the scattered fields, calculated for a single HR mirror at  $u_\text{cav}$.}
\label{Fig:SingleMirror}
\end{figure}

To further investigate the mechanisms that lead to this very high Q-factor, a continuous wave (CW) calculation of a single 5-period-MPC mirror has been performed, using a point source at $u_\text{cav}$.  Fig.~\ref{Fig:SingleMirror} shows (a) the absolute value and (b) the angle of the Poynting vector of the scattered field. By inspection of the figure, three effects are clearly evident: beam  confinement in transmission, strong decay of the optical power inside the mirror and focusing of the reflected beam.
In fact, on the right of the input interface, a narrow feature confined around $y=0$ interferes and decays inside the mirror. On the left side, a refocused beam forms and propagates back, resembling the focusing behaviour of a curved mirror and allowing the optical power to be steadily confined inside the cavity.
%%%%%%%%%%%%%%%%%%%%%%%%%%%%%%%%%%%%%%%%%%%

\begin{figure}[ht!]
\centering
\includegraphics[width=0.60\columnwidth]{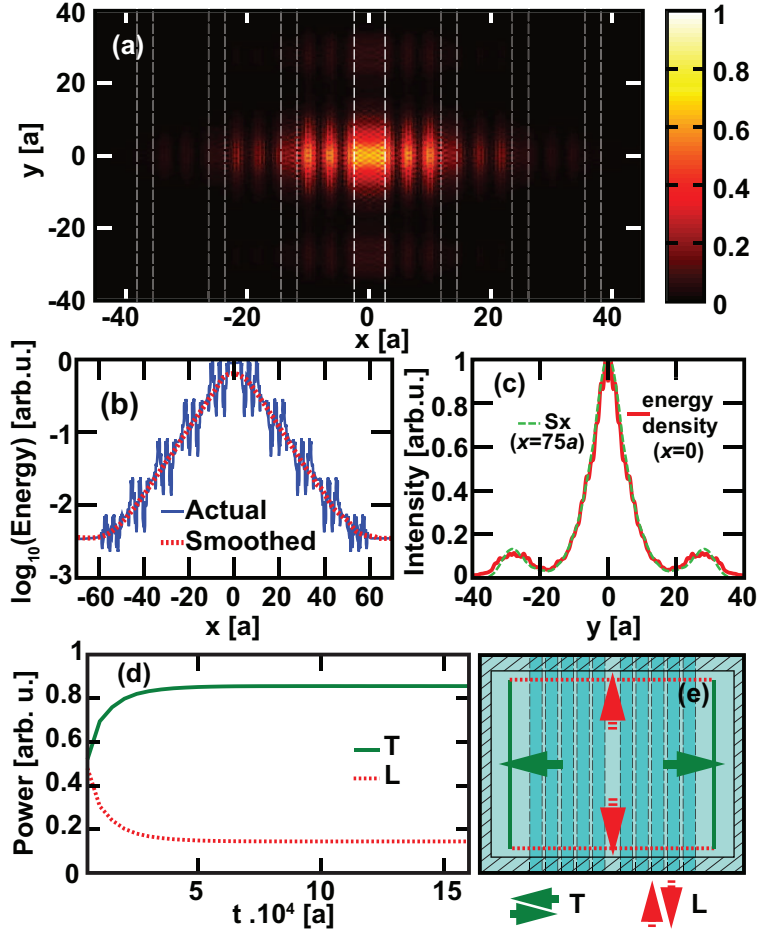}
\caption{(a) Total energy density inside the microcavity. (b) Logarithm of the density of the total energy integrated along the $y$-direction. (c) Profile $S_x$ of the $x$ component of the self-normalised Poynting vector of the output beam as a function of the $y$ coordinate, for $x=75\,a$ (green dashed curve); profile of the self-normalised energy density as a function of the $y$ coordinate, for $x=0$ (red solid curve). (d) Transmittance (green solid curve) and losses (red dotted curve) as a function of the normalised time. (e) Sketch of the calculation.}
\label{Fig:CWpanel}
\end{figure}

Similar CW simulation with a point source inside the defect is also performed on the whole cavity for the best configuration. Fig.~\ref{Fig:CWpanel}(a) depicts the energy density profile inside the cavity. On the $x$ direction, the energy density is confined in the vicinity of the defect, showing an exponential decay as the field penetrates into the mirrors. Fig.~\ref{Fig:CWpanel}(b) shows the logarithm of the self-normalised energy density, integrated along the $y$ direction, as a function of the $x$ coordinate; the superposed red dotted line represents the same quantity smoothed over one mesoscopic period. Thus, this total energy shows an exponential trend inside the mirrors with a decay constant of $-0.0459\,a^{-1} \pm 0.0002\,a^{-1}$ and a full width at half maximum along the $x$ direction of about $23.6\,a$. Along the $y$ direction, the energy density shows a gaussian-like profile with a waist of $9.1\,a$ (see Fig~\ref{Fig:CWpanel}(c)). It is worth pointing out that this waist does not depend on the overall size $L_\text{
H}$.
The mode confinement can be estimated by the normalised effective mode area $A_{\text{eff,n}}$, that can be defined as follows:
\begin{equation}
%A_{\text{eff,n}}=\frac{\iint_{-\infty }^{\infty }U(x,y)dxdy}{A_0U_{\text{max}}}, 
A_{\text{eff,n}}=\left(\iint_{-\infty }^{\infty }U(x,y)dxdy\right) /A_0U_{\text{max}}, 
\label{eq:Volume}
\end{equation}
where $U(x,y)$ is the total energy density, $U_{\text{max}}$ is its maximum value inside the computational domain and $A_0=(1 / n_{\text{eff}} u_\text{cav})^2$ is the diffraction-limited mode area. The calculated value is $A_{\text{eff,n}}=99.7$, which is around 2 order of magnitude greater than a typical H1 PhC cavity as reported in \cite{Ota2009}. Even though the mode area is large for quantum electrodynamic applications, it is suitable for biosensing and single or multiple isolated object detection. In fact, since 2D confinement can occur at any $y$ position along planar cavity with arbitrarily long vertical size ($L_\text{H}$), the overlapping probability with isolated objects is greatly enhanced. Furthermore, multiple objects can even overlap with different independent localized modes at different y position along the same planar cavity.

From a temporal point of view, as the source starts to radiate, the cavity begins to charge and the stored energy reaches $99.9\%$ of the saturation level after a normalised time equal to $t_s=11.4 \cdot 10^4\,a$. As shown in Fig.~\ref{Fig:CWpanel}(d), by considering a rectangle that embraces the computational cell without including the PML, the power transmitted through its vertical sides (green solid curve) and losses through its horizontal sides (red dotted curve) can be calculated as a function of the normalised time (see Fig.~\ref{Fig:CWpanel}(e)). These quantities are both normalised to  the total power flux $P_{S}$ of the point source without any surrounding structure. When the steady-state regime is reached, the transmittance and losses result higher than $85\%$ and lower that $15\%$, respectively.
These losses are ascribable to the point source spatial components that are scattered outside the mirrors without being refocused, mainly confined inside the defect. The transmitted power is carried out of the cavity by two symmetrical collimated beams that are formed thanks to the MSC inside the mirrors. Fig.~\ref{Fig:CWpanel}(c) shows the profile $S_x$ of the $x$ component of the Poynting vector along the $y$-direction, when $x=75\,a$. The beam shows a gaussian-like shape with the same waist as the total energy density profile inside the defect.

%%%%%%%%%%%%%
%Fig.~\ref{Fig:CWpanel}(d) also shows the Q factor as a function of the normalised time (black dashed curve, referring to the right vertical axis of the figure)  estimated by means of its general definition $Q=2 \pi u_\text{cav} {U_{\text{stored}}}/{P_{\text{S}}}$, where $U_{\text{stored}}$ is the total energy density integrated into the aforementioned rectangle. This curve reaches the saturation value of about 10100 after a normalised time equal to $t_s$. This value is in good agreement with that previously calculated.
%%%%%%%%%%%%

In conclusion, a novel stable planar microcavity based on HR MPCs has been designed. The analysis of the microcavity has been detailed by means of 2D-FDTD simulations. In particular, when the mirrors are only 5-period-long, the cavity shows a maximum Q factor of about 12840. Moreover, a Q factor of about 10000 is achievable for a wide range of cavity lengths, making the structure robust with respect to fabrication uncertainties. In the same range, the linear dependence of the resonant frequency on the cavity width eases the tunability of the structure. Furthermore, higher Q factors are expected using longer mirrors. Such high Q factors with small beam size and planar reflectors are achieved owing to  focusing effect arising in MPC, and cannot be reached using bulk layered materials.  In addition, our novel approach allows high Q with 2D confinement anywhere along the defect in an arbitrarily long planar cavity. This clearly improves overlap chances between isolated objects and the confined mode for field 
enhancement. Thus, possible device applications range from lasing in  active media, thanks to the beamforming behaviour of the structure, to field enhancement for biosensing, thanks to the high field localization and the possibility of 1D alignment instead of 2D alignment. To our knowledge,  this is the first example of stable planar cavity in literature.

The research has been conducted in the framework of the European Cooperation in Science and Technology (“COST”) Action MP0805. G. Cal\`o and V. Petruzzelli thank the Photonic Interconnect Technology for Chip Multiprocessing Architectures (“PHOTONICA”) project under the Fondo per gli Investimenti della Ricerca di Base 2008 (“FIRB”) program, funded by the Italian government and thanks to the facilities offered by the Apulia region laboratory project "Synthesis and characterization of new organic and nanostructured materials for electronics, photonics, and advanced technologies". M. G. thanks the U.S. Army International Technology Center Atlantic for 
financial support (W911NF-12-1-0292).

\bibliographystyle{ol}

\begin{thebibliography}{1}

\bibitem{Kosaka1999}
H.~Kosaka, T.~Kawashima, A.~Tomita, M.~Notomi, T.~Tamamura, T.~Sato, and
  S.~Kawakami, Applied Physics Letters \textbf{74}, 1212 (1999).

\bibitem{Chigrin2002}
D.~N. Chigrin, S.~Enoch, C.~M. Sotomayor-Torres, and G.~Tayeb, in
  \emph{Symposium on Integrated Optoelectronic Devices} (ISOP, 2002), pp. 63--72.

\bibitem{Chuang2010}
Y.~Chuang and T.~Suleski, Journal of Optics \textbf{12}, 035102 (2010).

\bibitem{Shen2011}
X.~Shen, T.~J. Cui, and J.~Ye, in \emph{Symposium on Photonics and Optoelectronics,
  2011} (IEEE, 2011), pp. 1--4.

\bibitem{Arlandis2012}
J.~Arlandis, E.~Centeno, R.~Polles, A.~Moreau, J.~Campos, O.~Gauthier-Lafaye,
  and A.~Monmayrant, Physical Review Letters \textbf{108}, 037401 (2012).

\bibitem{Magno2014}
G.~Magno, M.~Grande, A.~Monmayrant, F.~Lozes-Dupuy, O.~Gauthier-Lafaye, G.~Cal\'{o}, and V.~Petruzzelli, JOSA B, under press (2014).

\bibitem{Mocella2009}
V.~Mocella, S.~Cabrini, A.~Chang, P.~Dardano, L.~Moretti, I.~Rendina,
  D.~Olynick, B.~Harteneck, and S.~Dhuey, Physical Review Letters \textbf{102},
  133902 (2009).

\bibitem{Rakich2006}
P.~T. Rakich, M.~S. Dahlem, S.~Tandon, M.~S. Mihai~Ibanescu, G.~S. Petrich,
  J.~D. Joannopoulos, L.~A. Kolodziejski, and E.~P. Ippen, Nature Materials
  \textbf{5}, 93 (2006).

\bibitem{PTL08-20-24-2120-2122}
A.~Larrue, O.~Bouchard, A.~Monmayrant, O.~Gauthier-Lafaye, S.~Bonnefont,
  A.~Arnoult, P.~Dubreuil, and F.~Lozes-Dupuy, Photonics Technology Letters,
  IEEE \textbf{20}, 2120 (2008).

\bibitem{Oskooi2010687}
A.~F. Oskooi, D.~Roundy, M.~Ibanescu, P.~Bermel, J.~Joannopoulos, and S.~G.
  Johnson, Computer Physics Communications \textbf{181}, 687  (2010).

\bibitem{Ota2009}
Y.~Ota, M.~Shirane, M.~Nomura, N.~Kumagai, S.~Ishida, S.~Iwamoto, S.~Yorozu,
  and Y.~Arakawa, Applied Physics Letters \textbf{94}, 033102 (2009).

\end{thebibliography}

\end{document}